*TAPIR* enables high-throughput estimation and comparison of phylogenetic informativeness using locus-specific substitution models

Brant C. Faircloth[1*], Jonathan Chang[1], Michael E. Alfaro[1]

[1] Department of Ecology and Evolutionary Biology, University of California, Los Angeles, CA, 90095 USA

* To whom correspondence should be addressed

**Running Head**: Rapidly estimating phylogenetic informativeness

**Abstract**

Massively parallel DNA sequencing techniques are rapidly changing the dynamics of phylogenetic study design by exponentially increasing the discovery of phylogenetically useful loci. This increase in the number of phylogenetic markers potentially provides researchers the opportunity to select subsets of loci best-addressing particular phylogenetic hypotheses based on objective measures of performance over different time scales. Investigators may also want to determine the power of particular phylogenetic markers relative to each other. However, currently available tools are designed to evaluate a small number of markers and are not well-suited to screening hundreds or thousands of candidate loci across the genome. *TAPIR* is an alternative implementation of Townsend's estimate of phylogenetic informativeness (PI) that enables rapid estimation and summary of PI when applied to data sets containing hundreds to thousands of candidate, phylogenetically informative loci.

**Availability and Implementation**: *TAPIR* is written in Python, supported on OSX and linux, and distributed under a BSD-style license at: http://www.github.com/faircloth-lab/tapir/.

**Contact**: brant@faircloth-lab.org

**Supplemental information**: N/A

**1 Introduction**

Several barriers preclude the optimal design of phylogenetic studies. These include the lack of large numbers of genetic markers useable across the breadth of taxa under study [1] and the means of selecting the optimal set or subset(s) of markers to resolve phylogenetic hypotheses related to those taxa [2, 3]. The availability of genome sequences for comparative analysis [4] and marker design [5-7], new techniques for data collection [8, 9], and continuing advances in massively parallel DNA sequencing [10] are rapidly increasing the number of markers useful across broad taxonomic extents to overcome the first barrier. Quantitative methods of estimating the information content or informativeness of candidate loci are addressing the second.

Townsend [3] proposed an algorithm to enable the computation of phylogenetic informativeness (PI) at discrete time periods and across spans of time (epochs). Here, we report an implementation of Townsend's algorithm, *TAPIR* (Tally Approximations of Phylogenetic Informativeness Rapidly), suited to high-throughput analysis and comparison of large (> 100 loci) data sets. *TAPIR* expands on the capabilities of the PhyDesign [11] web application in several ways. First, *TAPIR* selects the best-fitting, finite-sites substitution model for each locus prior to inputting the computed base frequencies and estimated substitution rate matrix for each locus to the site rate and phylogenetic informativeness estimation procedure. This allows subsequent phylogenetic informativeness measures to incorporate more realistic models of locus-specific substitution. Second, *TAPIR* uses a parallel processing approach to estimate substitution models, site rates, and phylogenetic informativeness for large datasets (>100 loci) datasets reasonably quickly. Third, *TAPIR* enables rapid re-analysis of data from intermediate results stored in a structured format (JSON). Fourth, *TAPIR* collects results across loci in a SQL database, easing data summary and subsequent comparison and analysis of data sets. Finally, *TAPIR* provides helper scripts to facilitate visualization of computational results and comparative analyses of alternative data sets.

**2 Approach**

We wrote *TAPIR* in Python, taking advantage of the fast array operations provided by the *NUMPY* (http://numpy.scipy.org) and *SCIPY* (http://scipy.org) libraries, tree handling using *DENDROPY* (http://packages.python.org/dendropy/), and *SQLITE3* for data storage and retrieval. *TAPIR* also depends on *HYPHY* [12]. Briefly, *TAPIR* takes as input a dated tree, a folder of nexus-formatted alignments containing the taxa in the dated tree, a list of discrete time points for which to compute the net phylogenetic informativeness ($D_{PI}$), a list of intervals over which to compute the net phylogenetic informativeness ($I_{PI}$), and an output folder for results storage. After starting a run, *TAPIR* generates an array of discrete times spanning the depth of the dated tree

$$T_{Tree} = \begin{bmatrix} T_1 \\ T_2 \\ ... \\ T_{n-1} \\ T_n \end{bmatrix}$$

and scales the branch lengths of the input tree to fall within the interval $(0, 100]$ using a correction factor ($C$). Then, the program feeds each alignment to a *HYPHY* sub-process that computes the best-fitting, finite-sites substitution model for the alignment, estimates the site rates across each alignment given the best-fitting substitution model, scales the site rates by $C$, creates an array of corrected site rates

$$S_C = [\lambda_1 \quad \lambda_2 \quad ... \quad \lambda_{n-1} \quad \lambda_n]$$

and outputs raw and corrected site rates ($S_C$) to a JSON-formatted file in a user-selected output folder. To minimize the introduction of sites with poorly estimated rates to the analysis, *TAPIR* masks (as null values) $S_C$ at positions where there are fewer than the user-supplied (default = 3) number of sites. To compute net phylogenetic informativeness for discrete time periods, *TAPIR* inputs the two arrays of data, $T_{Tree}$ and $S_C$, to Townsend's [3] equation:

$$\rho(T; \lambda) = 16\lambda_i^2 T e^{-4\lambda_i T}$$

and computes results element-wise:

$$\begin{bmatrix} 16\lambda_1^2 T_1 e^{-4\lambda_1 T_1} & ... & 16\lambda_n^2 T_1 e^{-4\lambda_n T_1} \\ ... & ... & ... \\ 16\lambda_1^2 T_n e^{-4\lambda_1 T_n} & ... & 16\lambda_1^2 T_n e^{-4\lambda_1 T_n} \end{bmatrix}$$

*TAPIR* sums across the axes of the resulting array to compute net informativeness for each time in $T_{Tree}$ and returns requested values to the user by reindexing the array. To compute phylogenetic informativeness over intervals, *TAPIR* iterates over user-defined epochs $[start, end]$ and uses scipy to vectorize the integral computation of (eqn. 1) over $[start, end]$

$$\int_{start}^{end} \rho(T; \lambda) dT$$

using the QUADPACK algorithm[13].

*TAPIR* writes results for all computations of PI to an *SQLITE* database indexed by locus name. If running on a platform having multiple compute cores, *TAPIR* divides the number of loci to be analyzed into subsets of roughly equal size and processes subsets of data in parallel using *n-1* compute cores. Users can rapidly re-process site-rate data to re-compute $D_{PI}$ and $I_{PI}$ at different times or across different intervals by passing the output folder containing $S_C$ as input to *TAPIR* along with a command-line flag. We provide helper scripts within the *TAPIR* package that support graphical presentation and comparison of results from different marker sets using RPY2 (http://rpy.sourceforge.net/). Users can generate more complex figures by connecting a statistical/graphics package (i.e., R) to the results store in the sqlite database.

To illustrate the graphical outputs of the program and summarize the amount of time required to process loci, we selected three data sets containing 20, 183, and 917 nuclear loci [14, 15], each drawn from the same 17 (Supplementary Table 1), genome-enabled mammals, and we estimated net PI ($D_{PI}$ and $I_{PI}$) across 5 intervals intersecting each node of the same dated tree (Supplementary Fig 1). We ran all computations using an Apple Mac Pro workstation (dual, quad-core Intel Xeon at 3 Ghz) having 24 GB of RAM, and we plotted the resulting PI values ($I_{PI}$) for the data set containing 20 nuclear loci across each of 10 time intervals (Supplementary Fig 2). We used the comparative plotting functions of *TAPIR* to contrast the mean (Fig 1A) and net (Fig 1B) PI of the 20 most informative loci from each data set during select time intervals. We also plot the run time (Supplementary Figure 4) required for *TAPIR* to process each data set.

**ACKNOWLEDGEMENTS**

BCF thanks SP Hubbell, PA Gowaty, RT Brumfield, TC Glenn, NG Crawford, and JE McCormack. We thank Francesc Lopez-Giraldez and Jeffrey Townsend for providing us with a copy of their web-application source code and helpful discussion. An Amazon Web Services Research Grant to BCF and NSF grants DEB 6861953 and DEB 6701648 to MEA partially supported this work.

**REFERENCES**

1. Murphy WJ, Eizirik E, Johnson WE, Zhang YP, Ryder OA, O'Brien SJ: **Molecular phylogenetics and the origins of placental mammals.** *Nature* 2001, **409**:614–618.

2. Goldman N: **Phylogenetic information and experimental design in molecular systematics.** *Proc Biol Sci* 1998, **265**:1779–1786.

3. Townsend JP: **Profiling phylogenetic informativeness.** *Systematic Biol.* 2007, **56**:222–231.

4. Miller W, Rosenbloom K, Hardison RC, Hou M, Taylor J, Raney B, Burhans R, King DC, Baertsch R, Blankenberg D, Pond SLK, Nekrutenko A, Giardine B, Harris RS, Diekhans STM, Diekhans M, Pringle TH, Murphy WJ, Lesk A, Weinstock GM, Lindblad-Toh K, Gibbs RA, Lander ES, Siepel A, Haussler D, Kent WJ: **28-way vertebrate alignment and conservation track in the UCSC Genome Browser**. *Genome Res* 2007, **17**:1797–1808.

5. Fulton TM: **Identification, Analysis, and Utilization of Conserved Ortholog Set Markers for Comparative Genomics in Higher Plants**. *The Plant Cell* 2002, **14**:1457–1467.

6. Ranwez V, Delsuc F, Ranwez S, Belkhir K, Tilak M-K, Douzery EJ: **OrthoMaM: a database of orthologous genomic markers for placental mammal phylogenetics.** *BMC Evol. Biol.* 2007, **7**:241.

7. Faircloth BC, McCormack JE, Crawford NG, Harvey MG, Brumfield RT, Glenn TC: **Ultraconserved Elements Anchor Thousands of Genetic Markers for Target Enrichment Spanning**. *Systematic Biol.*, **In press**.

8. Mamanova L, Coffey AJ, Scott CE, Kozarewa I, Turner EH, Kumar A, Howard E, Shendure J, Turner DJ: **Target-enrichment strategies for next-generation sequencing.** *Nat Methods* 2010, **7**:111–118.

9. Gnirke A, Melnikov A, Maguire J, Rogov P, LeProust EM, Brockman W, Fennell T, Giannoukos G, Fisher S, Russ C, Gabriel S, Jaffe DB, Lander ES, Nusbaum C: **Solution hybrid selection with ultra-long oligonucleotides for massively parallel targeted sequencing**. *Nature Biotechnology* 2009, **27**:182–189.

10. Shendure J, Ji H: **Next-generation DNA sequencing.** *Nat Biotechnol* 2008, **26**:1135–1145.

11. López-Giráldez F, Townsend JP: **PhyDesign: an online application for profiling phylogenetic informativeness.** *BMC Evol. Biol.* 2011, **11**:152–.

12. Pond SLK, Frost SDW, Muse SV: **HyPhy: hypothesis testing using phylogenies.** *Bioinformatics* 2005, **21**:676–679.

13. Piessens R, de Doncker-Kapenga E, Uberhuber C: *QUADPACK: A Subroutine Package for Automatic Integration*. 1st edition. Berlin: Springer; 1983.

14. McCormack JE, Faircloth BC, Crawford NG, Gowaty PA, Brumfield RT, Glenn TC: **Ultraconserved Elements Are Novel Phylogenomic Markers that Resolve Placental Mammal Phylogeny when Combined with Species Tree Analysis.** *Genome Res* 2011.

15. Springer MS, Burk-Herrick A, Meredith R, Eizirik E, Teeling E, O'Brien SJ, Murphy WJ: **The adequacy of morphology for reconstructing the early history of placental mammals**. *Systematic Biol.* 2007, **56**:673–684.

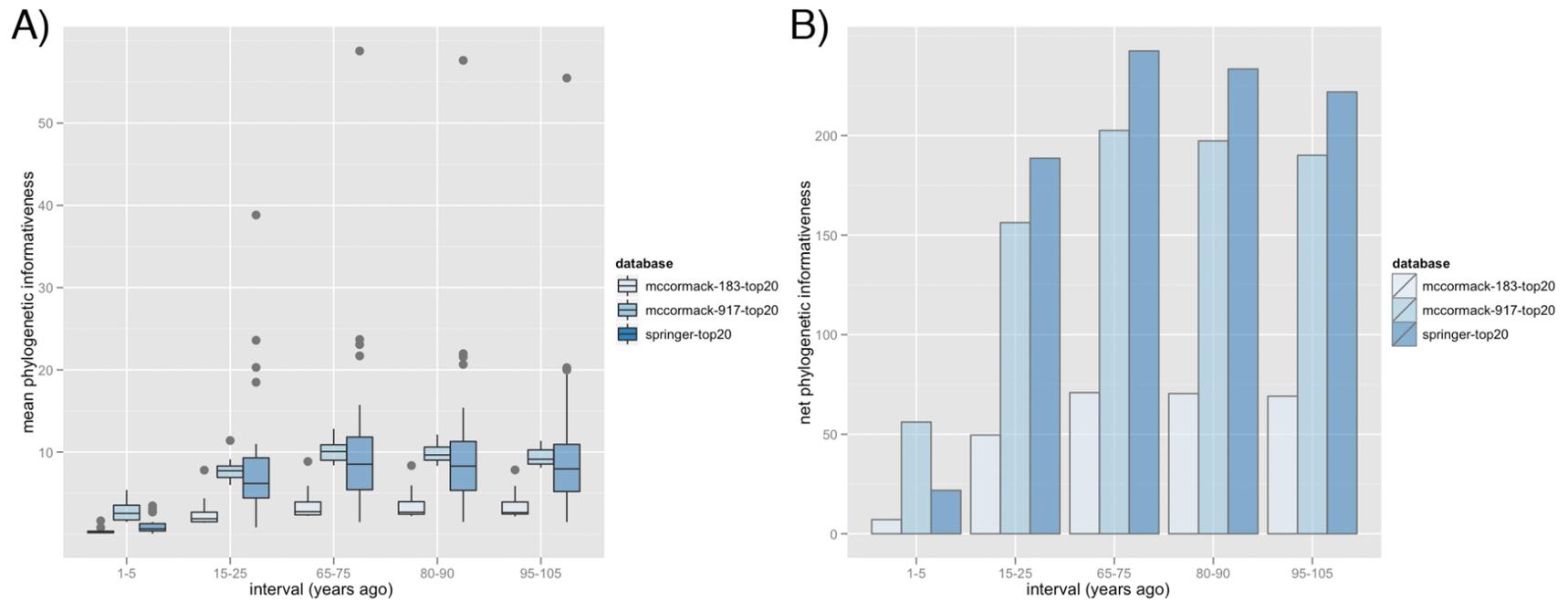

Figure 1. Sample plots illustrating a subset of the graphical outputs from *TAPIR* that allow the user to visualize and compare phylogenetic informativeness within and among datasets. Here, we compare (A) mean phylogenetic informativeness and (B) net phylogenetic informativeness of the 20 most informative loci across a subset of intervals (MYA) computed across three different data sets.

Supplementary Table 1. Genome sequences used to select nuclear loci for alignment and computation of phylogenetic informativeness and construction of the dated tree.

| Abbr | Scientific name | Taxon | Genome Build Version |
|---|---|---|---|
| anoCar2 | *Anolis carolinensis* | Anole | Broad release anoCar2.0 |
| taeGut1 | *Taeniopygia guttata* | Zebra finch | WUSTL release v3.2.4 |
| monDom5 | *Monodelphis domestica* | Opposum | Broad Release monDom5 |
| loxAfr3 | *Loxodonta africana* | Elephant | Broad release loxAfr3 |
| bosTau4 | *Bos taurus* | Cow | Baylor release 4.0 |
| canFam2 | *Canis familiaris* | Dog | Broad release canFam2 |
| equCab2 | *Equus caballus* | Horse | Broad release Equus2 |
| oryCun2 | *Oryctolagus cuniculus* | Rabbit | Broad release oryCun2 |
| cavPor3 | *Cavis porcellus* | Guinea pig | Broad release cavPor3 |
| mm9 | *Mus musculus* | Mouse | NCBI Build 37 |
| calJac3 | Callithrix jacchus | Marmoset | WUSTL release calJac3.2 |
| rheMac2 | *Macaca mulatta* | Macaque | Baylor release v.1.0 Mmul_051212 |
| ponAbe2 | *Pongo abelii* | Orangutan | WUSTL release Pongo_albelii-2.0.2 |
| gorGor3 | *Gorilla gorilla* | Gorilla | Wellcome Trust Sanger Institute release 57 |
| panTro2 | *Pan troglodytes* | Chimp | Chimp Genome Sequencing Consortium Build 2 version 1 |
| venter | *Homo sapiens* | Craig Venter genome | JCVI HuRef 1.0 |
| korean | *Homo sapiens* | Korean individual | KOBIC-KoreanGenome KOREF_20090224 |
| chinese | *Homo sapiens* | Chinese individual | BGI YH Genome |
| hg19 | *Homo sapiens* | Human reference | Genome Reference Consortium Human Reference 37 |

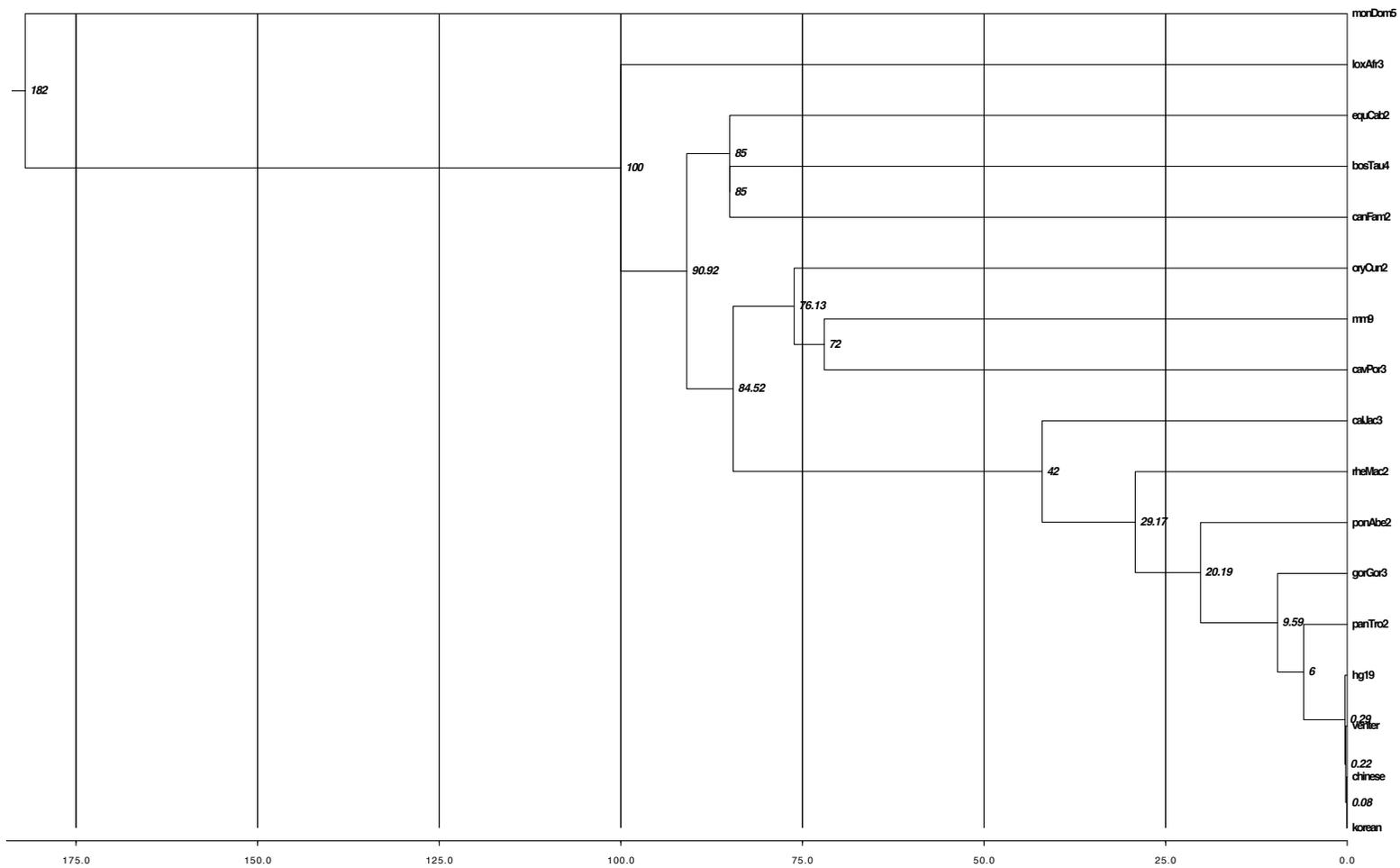

Supplementary Figure 1. Dated tree used for all computations of phylogenetic informativeness.

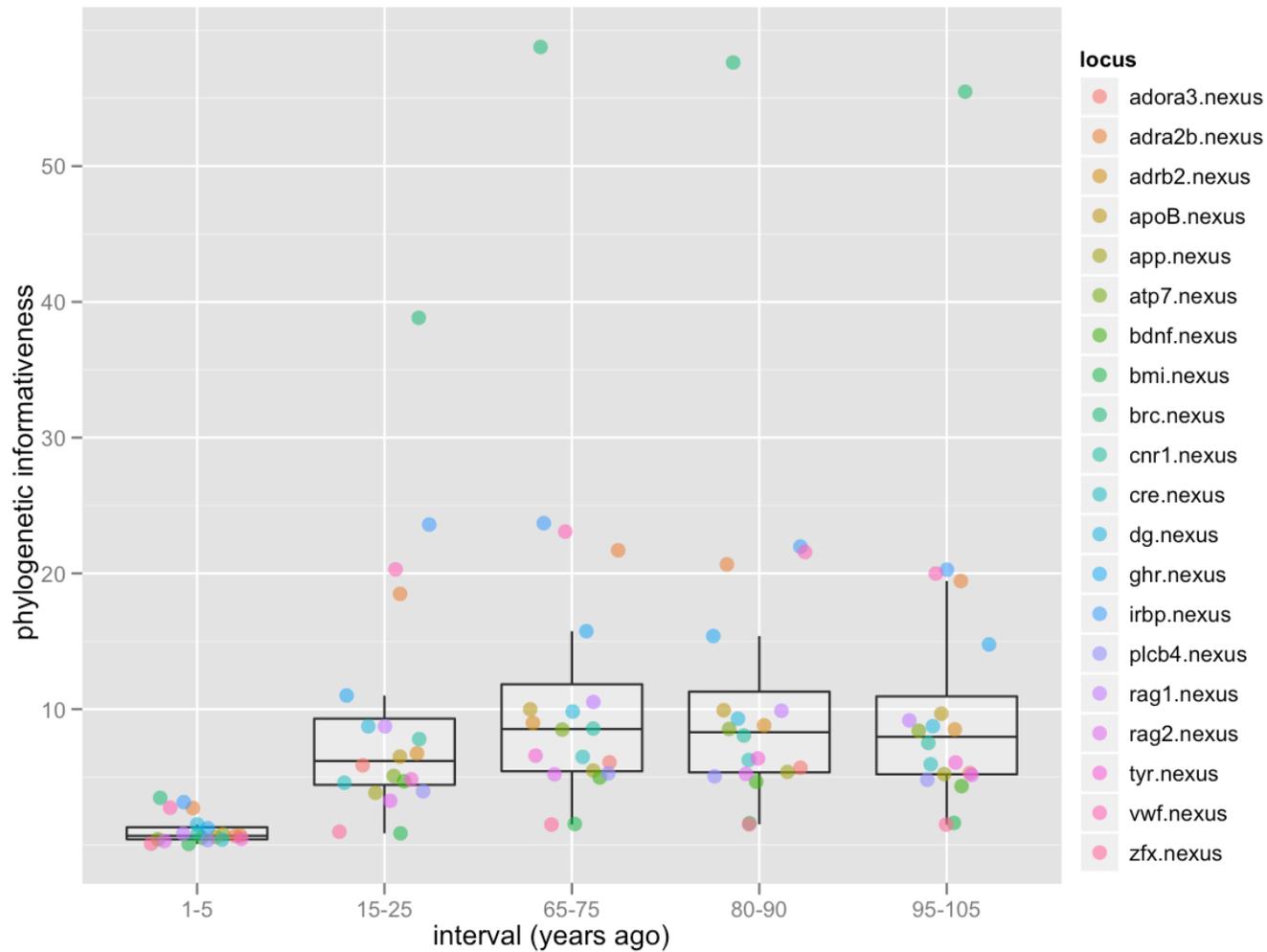

Supplementary Figure 2. Mean phylogenetic informativeness of commonly used nuclear loci [15] for resolving select nodes of the dated tree in Supplementary Figure 1.

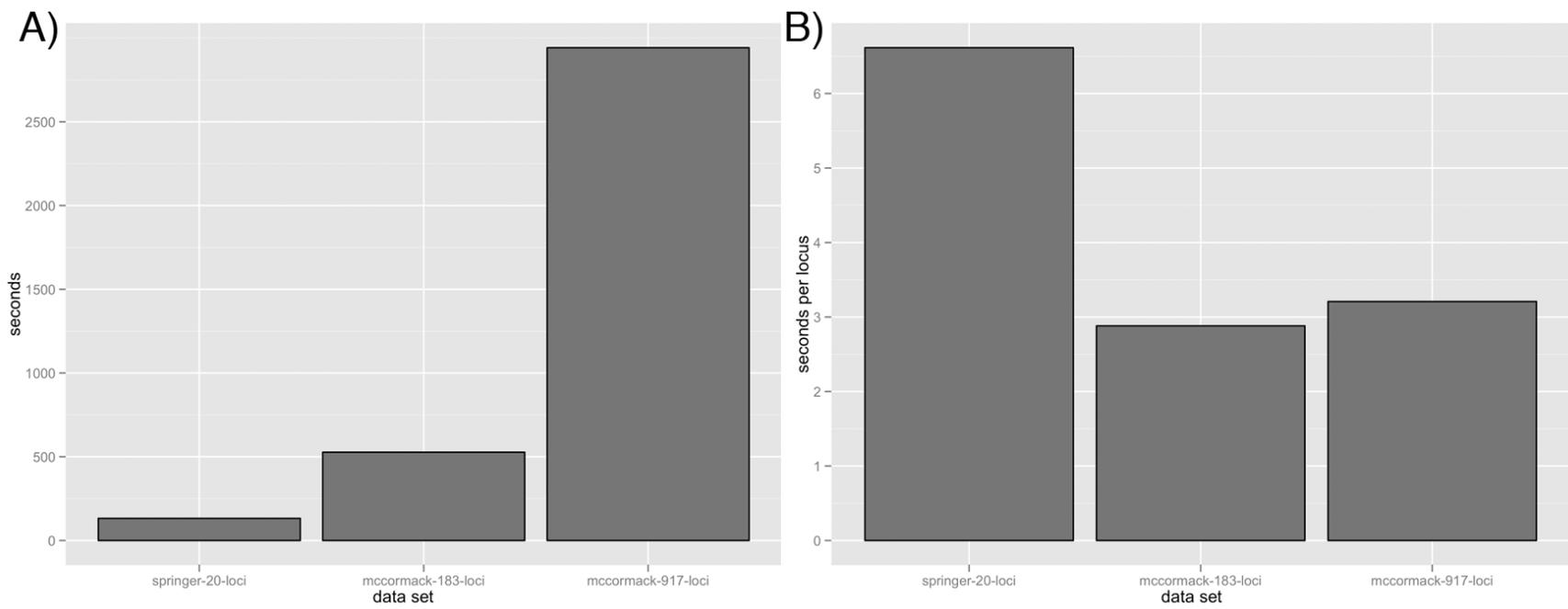

Supplementary Figure 3. Processing time for all data sets.